\documentclass[preprint,12pt]{elsarticle}

\usepackage{epsfig}

\usepackage{amssymb}
\journal{Nuclear Physics A}

\begin{document}

\begin{frontmatter}



\title{Understanding LHC jets in the light of RHIC data}


\author{Thorsten Renk}

\address{Department of Physics, P.O. Box 35, FI-40014 University of Jyv\"askyl\"a, Finland and \\Helsinki Institute of Physics, P.O. Box 64, FI-00014 University of Helsinki, Finland}

\begin{abstract}
Hard probes are a cornerstone in the ongoing program to determine the properties of hot and dense QCD matter as created in ultrarelativistic heavy ion collisions. LHC measurements have so far resulted in a wealth of high $P_T$ data, opening new kinematic windows with high statistics. Yet on first glance, several observations are counter-intuitive and seem to contradict results from the RHIC high $P_T$ program. This calls for a combined analysis of high $P_T$ hadrons and reconstructed jets  at RHIC and LHC in a unified framework testing a large number of theoretical models for both medium evolution and shower medium interactions against the systematics of the data. A consistent picture of shower-medium interaction emerges from this analysis which explains where and why results appear counter-intuitive.

\end{abstract}

\begin{keyword}
jet quenching \sep quark gluon plasma


\end{keyword}

\end{frontmatter}

% \linenumbers

\section{What did we know about energy loss?}
\label{}

Hard  perturbative Quantum Chromodynamics (pQCD) processes taking place along with the creation of soft bulk matter in heavy-ion collisions are a  calculable source of high $p_T$ partons embedded in the medium. The medium modification of high $P_T$ observables through the final state interaction of  parton showers evolving in the expanding medium thus carries information about both global (i.e. geometry) and local medium properties (i.e. degrees of freedom determining the detailed physics of parton-medium interaction). 

The medium modification affects the whole parton shower, however for observables only sensitive to the leading shower partons, the modification can be cast into the somewhat simpler form of leading parton energy loss (see e.g. \cite{SysJet}), which for historical reasons is the term often used and discussed.

Hard probes at RHIC have established a few key properties of energy loss: 1) energy loss $\Delta E$ is not fractional, i.e. can not be written in a form $\Delta E \sim zE$ with $E$ the original parton energy. Assuming fractional energy loss leads to a decrease of the nuclear suppression factor $R_{AA}$ with increasing $P_T$ which is opposite to the trend observed in the data \cite{gamma-h}. 2) energy loss is not dominated by incoherent processes. Making this assumption predicts a linear dependence of energy loss on pathlength $L$ which is not supported by the measured dependence of the suppression factor on the angle of outgoing hadrons with the recation plane \cite{ElPh,EMC}. This constrains an incoherent component to the total energy loss  from above to about 10\%. 3) lost energy does not all appear as soft medium-induced gluon radiation. Making this assumption leads to a discrepancy with the measured dihadron correlation suppression factor $I_{AA}$ \cite{DihadronEl} and allows to constrain energy deposition into medium excitation (for instance via elastic collisions) from below to about 10\%, in nice agreement with the constraint from pathlength. 

Note that detailed modelling of the medium is absolutely crucial to obtain reliable results from a systematic multi-observable analysis --- serveral findings change even qualitatively if simplifying assumptions such as power law parton spectra are made.

\section{What was expected for LHC?}

From these findings, important expectations for fully reconstructed jet observables can be formed. For instance, in a vacuum shower, the longitudinal momentum distribution (i.e. the fragmentation function (FF)) is created in a series of partonic $1 \rightarrow 2$ splittings of a parent parton $i$ into daughters $j,k $ with decreasing virtuality scale via the splitting kernels $P_{i\rightarrow j,k}(z)$ where $z = E_j/E_i$. These splitting kernels are scale invariant, and as a result the fragmentation function is self-similar. 
If $\Delta E \sim E$, the medium-modified FF (MMFF) could be generated by modified splitting kernels $P'_{i\rightarrow j,k}(z)$ and yield a self-similar result with a different form than in vacuum. However, since energy loss can not be cast into this form, a different assumption is needed and the next natural scale at which the MMFF is modified is few times the medium temperature $T$, i.e. jets should exhibit strong modifications only below a constant energy (not constant $z$) fairly independent of jet energy \cite{Asymptotics}.

\begin{figure}[htb]
\begin{center}
\epsfig{file=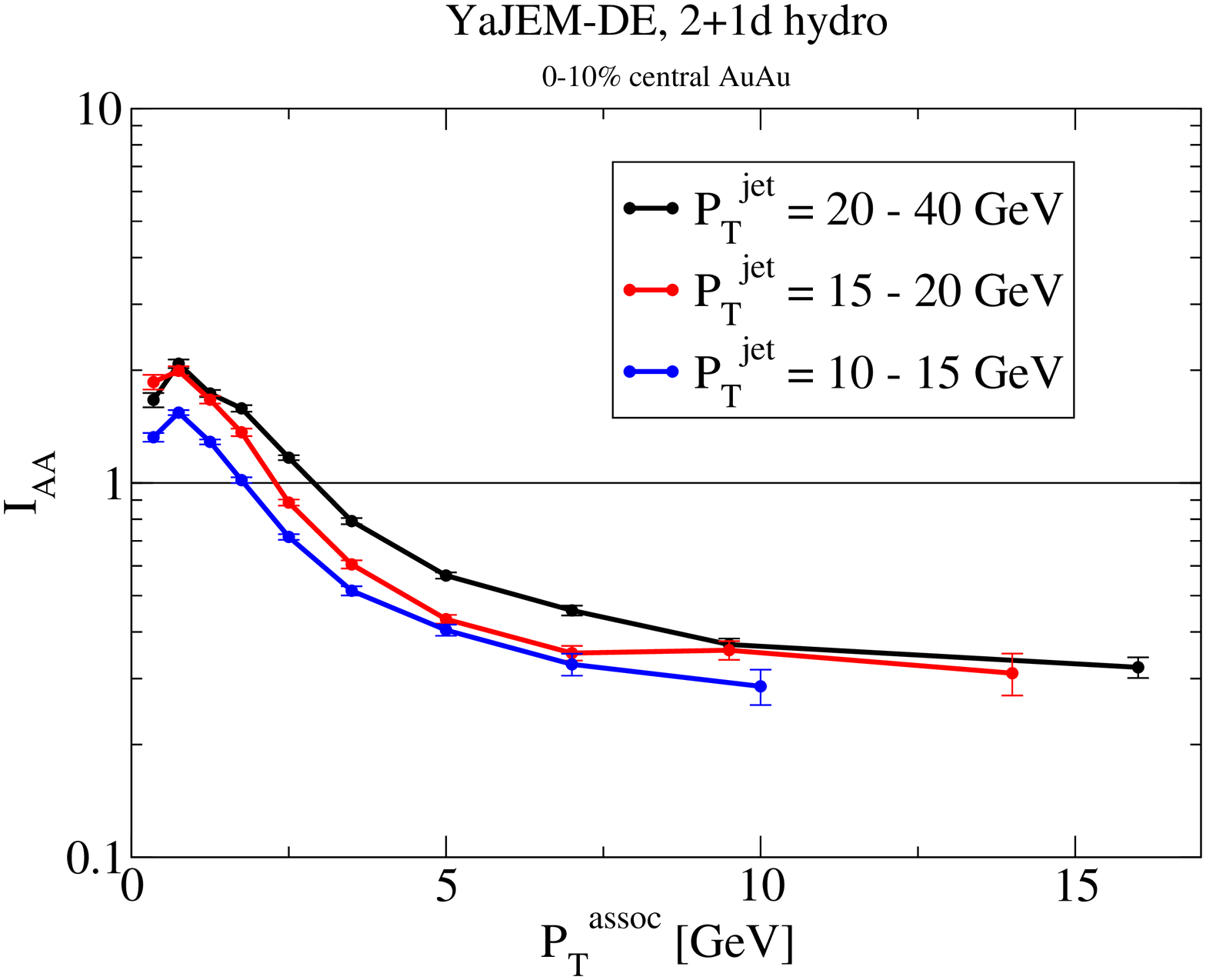, width=6cm}\epsfig{file=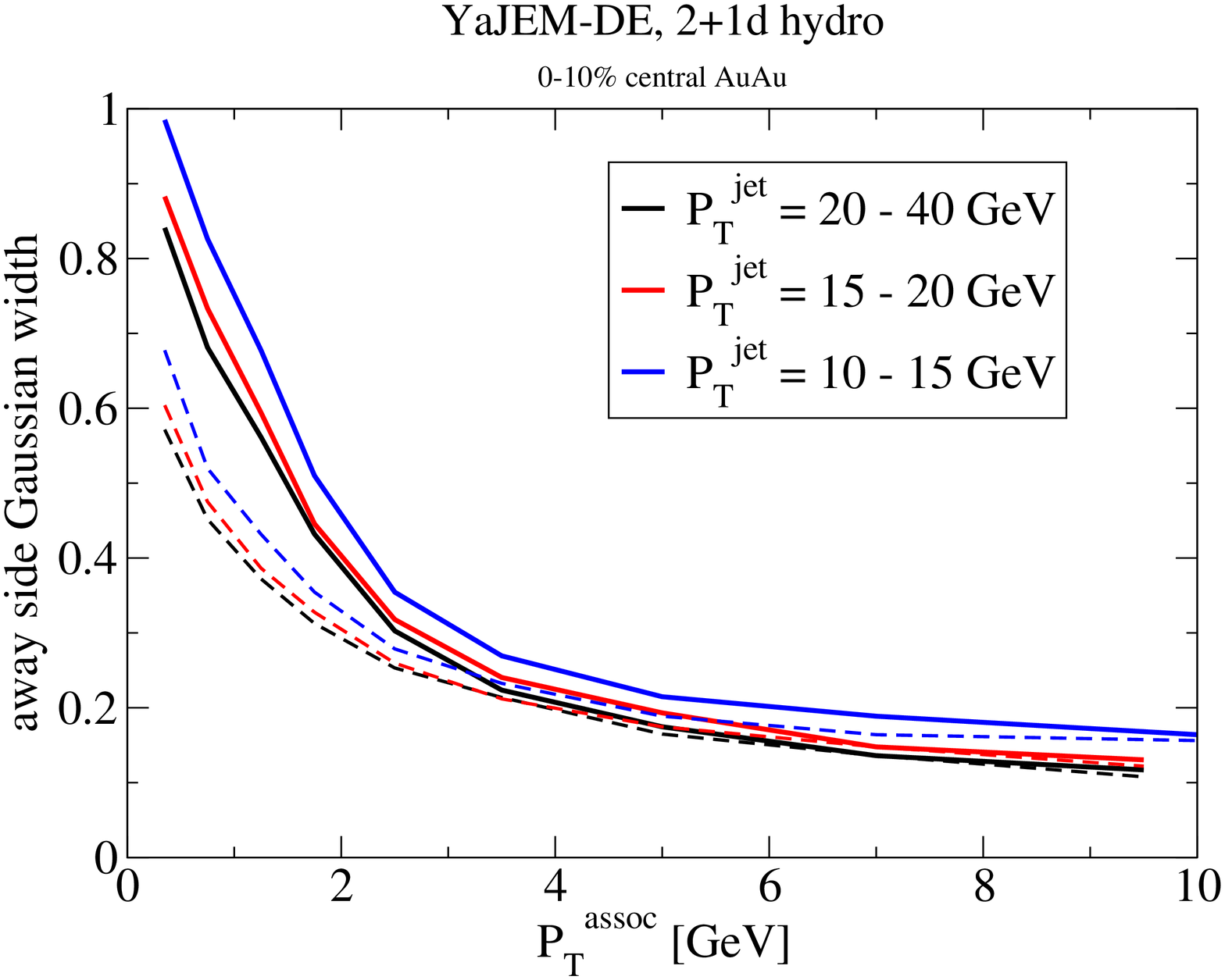, width=6cm}
\end{center}
\caption{\label{F-1}Left: Away side $I_{AA}$ as a function of associated $P_T$ for jet-h correlations at different trigger energies in central 200 AGeV Au-Au collisions. Right: Away side Gaussian width of the correlation peak for the vacuum case (dashed) and the medium case (solid) in the same calculation.}
\end{figure}

Computations with the in-medium shower evolution code YaJEM \cite{YaJEM} (in Fig.~\ref{F-1} for jet-h correlations at RHIC) show that the MMFF in such a scenario is flat (but suppressed) above a scale of $\sim 3$ GeV and strongly increases below. The same is true for the transverse momentum distribution which broadens much beyond the vacuum distribution at the same momentum scale. This behaviour has been observed by the STAR and the CMS collaboration \cite{FFexp}.

The underlying physics picture of jet quenching is here a multi-step process in which first the medium alters the hard parton kinematics slightly, leading to medium-induced soft gluon emission. This induced radiation is however collinear (and would not lead to jet quenching), but the re-scattering of the soft component leads to its quick thermalization and distribution to wide angles, largely independent of specific physics. Such a picture in which jet properties are only modified below a given momentum scale has been expected using YaJEM in \cite{YaJEM-jets}.

\section{Is this picture confirmed by LHC data?}

In all cases studied so far, the underlying physics picture has been shown to be in fair agreement with LHC data. This includes the measured dijet imbalance as a function of cone radius and jet energy and jet $R_{CP}$ \cite{LHC_AJ} as well as single hadron $R_{AA}$ \cite{LHC_RAA} and h-jet correlations as measured by ALICE \cite{h-jet}. More tests are currently work in progress.
The resulting constraints from various observables to different model assumptions can be cast into the form of a table (for details, see \cite{SysJet}):

\vspace{1ex}

{\footnotesize
\begin{tabular}{|l|cccccc|}
\hline
&  $R_{AA}^{RHIC}(\phi)$ & $R_{AA}^{LHC}(P_T)$ & $I_{AA}^{RHIC}$ & $I_{AA}^{LHC}$ &$A_J^{LHC}$& $A_J^{LHC}(E)$\\
\hline 
elastic &  {\bf fails!} &  {works} &  {\bf fails!} &  {fails}&  {works} &  {fails}\\
ASW &    {works} &  {fails} &  {marginal}  &  {works} & N/A & N/A\\
AdS &    {works} &  {\bf fails!} &  {marginal}  &  {works} & N/A & N/A\\
YaJEM &  {fails} &  {fails} &  {fails} &  {fails} &  {works} &  {works}\\
YaJEM-D &  {works} &  {works} &  {marginal} &  {marginal} & {works} &  {works} \\
YaJEM-DE &  {works} &  {works} &  {works} &  {works}& {works} &  {works}\\
\hline
\end{tabular}
}

From this, it becomes readily apparent that currently the main constraining information is already found from RHIC single hadron and correlation observables and that LHC data largely confirms the picture. The resulting physics picture is thus largely medium-induced pQCD radiation with a small component of energy depleted directly into medium degrees of freedom, with no sign of 'exotic' behaviour (such as suggested by AdS/CFT calculations). This indicates a shift from 'new ideas' towards quantitative understanding.



%% References


%\bibliographystyle{model2-num-names}



\end{document}